\documentclass[useAMS,usenatbib]{mn2e}
\usepackage{epsfig}
\usepackage{graphicx}
\usepackage{amssymb}
\usepackage{color}

\title[The location of the $\gamma$-ray emission region]{On the location of the $\gamma$-ray emission region for 21
flat spectrum radio quasars with quasi-simultaneous observations}
\author[G. Cao and J. C. Wang]{Gang Cao\thanks{E-mail: gcao@ynao.ac.cn} and Jiancheng Wang\thanks{E-mail: jcwang@ynao.ac.cn}\\
Yunnan Observatory, Chinese Academy of Sciences, Kunming 650011, China.}

\begin{document}
\pagerange{\pageref{firstpage}--\pageref{lastpage}} \pubyear{2013}

\maketitle

\label{firstpage}

%%%%%%%%%%%%%%%%%%%%%%%%%%%%%%%%%%%%%%%%%%%%%%%%%%%%%%%%%
\begin{abstract}
We try to infer the location of the GeV emission region for 21 flat spectrum radio quasars (FSRQs) with quasi-simultaneous spectral energy distributions (SEDs), in which the SEDs of 21 FSRQs are reproduced by the one-zone leptonic model including the synchrotron-self Compton (SSC) and external Compton (EC) processes. We suggest that the X-ray emission could be produced by the SSC process and the GeV emission could come from the EC process. The EC emission could originate from the inverse Compton (IC) scattering of photons from the broad line region (BLR) and accretion disk  or dust torus by the same electron population, which mainly depend on the location of the $\gamma$-ray emission region. We propose a method to constrain the location of the GeV emission region based on the spectral shapes. When the GeV emission is located within the BLR, the IC scattering could occur at the Klein-Nishina regime and produce a broken/steep spectrum in the GeV energy band. When the GeV emission is produced outside the BLR, the IC scattering could take place at the Thomson regime and the GeV spectrum  would have the same spectral index as the optical-infrared spectrum. We infer that the location of the GeV emission region is inside the BLR for 5 FSRQs and  beyond the BLR for 16 FSRQs. Our results show that the ratio of the magnetic field and electron energy density is close to equipartition condition for 21 FSRQs.

\end{abstract}


%--------------------------------------------------------------------------------------------------------------------------------------------------------
\section{Introduction}

Blazars are the most extreme subclass of Active Galactic Nuclei (AGNs). Their radiation is considered to originate in a relativistic jet oriented at a small angle with respect to the line of the sight. They show strong and rapid variability, high and variable polarization, and superluminal motions, etc. There are two subclasses of blazars, namely flat spectrum radio quasars (FSRQs) and BL Lac objects. BL Lacs  have weak or absent optical emission lines, while FSRQs usually show strong broad emission lines. The spectral energy distributions (SEDs) of blazars are dominated by non-thermal emission and consist of two distinct, broad components. It is generally agreed that the low component of their SEDs is produced by synchrotron emission from the relativistic electrons, whereas the origin of high component is still a matter of debate.  There are two classes of models to explain high energy emission: leptonic model and hadronic model. In the leptonic model, the high energy emission is produced by inverse Compton (IC) scattering of the soft photon fields (e.g., B\"{o}ttcher 2007a), in which soft photons are the synchrotron photons within the jet (the SSC process; Maraschi et al. 1992; Bloom and Marscher 1996; Zhang et al. 2012 ) or the photons external to the jet (the EC process). These external photons are the UV accretion disk photons \citep{der93} or the accretion disk photons reprocessed by broad line region clouds \citep{sik94} or the infrared (IR) photons from the dust torus \citep{bla00}. In the hadronic model, the high energy emission originates from proton synchrotron or photon-hadronic interaction \citep{man93,muc03,bot09}.

Since the emitting region is unresolved, the location of GeV $\gamma$-ray emission region is still an open issue.  Is the location within the BLR at sub-pc scale or beyond the BLR at pc scales where the torus IR emission dominates? \citet{tav10} argued that the emission region is within the BLR based on the rapid $\gamma$-ray variability. \citet{pou10} explained the Fermi GeV spectral break as the result of pair absorption on the He Lyman recombination continuum and lines from the BLR, implying that the GeV emission is within the BLR. However, the multiwavelength variability and very long baseline array images(VLBI) in the jets of several blazars placed the GeV emission at few pc distance from  the black hole \citep{mar10,jor10,agu11}. \citet{sik09} suggested that the $\gamma$-rays are produced by Compton scattering of the IR photon from dust torus due to the lack of bulk-Compton and Klein-Nishina (KN) effects.
\citet{sah12} reproduced the SEDs of 3C 279 using the one-zone leptonic model, they suggested that the observed very high energy (VHE) emission from this source supports an EC origin of the seed photons from the dust torus.
Recently, \citet{liu11} proposed a method to constrain the $\gamma$-ray emission position by the observed time lags of $\gamma$-rays relative to the broad emission. \citet{dot12} proposed a diagnostic test for the location of the GeV emission based on the energy dependence of the GeV variability. When the GeV emission locates with the BLR, the IC scattering could take place at the KN regime, in which the electron cooling time almost becomes energy independent and the variation of the $\gamma$-ray emission is expected to be achromatic. When the emission site locates outside the BLR, the IC scattering could take place in the Thomson regime, in which the GeV variability caused by electron cooling is energy dependent and faster at higher energy. But the above methods only apply to a small number of bright Fermi blazars.

We propose a diagnostic test for the location of the GeV emission based on the spectral shape. The basic view is the following as: when the GeV emission is outside the BLR, the IC scattering takes place at the Thomson regime, in which the GeV spectral index is the same as the synchrotron index; when the GeV emission is within the BLR, the IC scattering takes place at the KN regime and produces a broken spectrum. We reproduced the simultaneous SEDs of 21 blazars given by \citet{abd10b} to test the location of the GeV emission.

In Section 2, we describe the GeV spectral features. In Section 3 we present the method to reproduce the SEDs using a one-zone leptonic model with the SSC  and EC processes. In Section 4 we apply the method to 21 blazars. Throughout this paper, we adopt the cosmological
parameters of  $H_0 = 70 km \cdot s^{-1} \cdot Mpc^{-1}$, $\Omega_m = 0.3$, $\Omega_{\Lambda} = 0.7$.

\section{The GeV spectral features}
The Large Area Telescope LAT onboard the Fermi satellite provides unprecedented sensitivity in the $\gamma$-rays (20 MeV$-$300 GeV), and has observed 106 bright $\gamma$-ray sources with high confidence in the first three months of operation \citep{abd09c}, which is called the LAT bright AGN sample (LBAS). The LBAS contains 58 FSRQs, 42 BL Lac objects, 2 radio galaxies and 4 blazars of unknown type. \citet{abd10b} gave the quasi-simultaneous SEDs of 48 blazars with the Fermi, Swift and other observations. Recently, \citet{gio12} presented the simultaneous SEDs of 105 blazars, in which the synchrotron peak of FSRQs is at $\nu_s\sim10^{13}Hz$, while the IC peak is at $\nu_c\sim10^{22}Hz$. The simultaneous SEDs are important for us to understand the physical properties in the jet of blazars. We collect the quasi-simultaneous SEDs of 21 FSRQs which are taken from \citet{abd10b}.

We use a power-law model to fit  the optical and GeV spectra. If the GeV spectrum shows a break, we use a broken power-law model to fit the GeV spectrum. The fitting results are listed in table 1.
We classify them as two classes based on the spectral shapes.
First class includes 5 sources, in which the optical spectra are steeper than the GeV spectra.
For the 3 sources (4C 66.20, PKS 0454-234, and PKS 1502+106),
their GeV spectrum may show a break at a few GeV energy despite the large errors.
For  PKS 1454-354, the GeV spectrum of this source may show a break at $\sim1$ GeV from our fit, which is near the transition energy from the Thomson regime to KN regime. Furthermore, we also can't exclude that the broken energy is about at a few GeV energy due to the few data points and the large statistical errors.
For S4 0133+47, the GeV spectrum of this source may be consistent with a power-law, and show a break at a few GeV energy.
However, we can't identify the broken energy because of the few GeV data.
Second class has 16 sources, in which the GeV spectra have a power-law shape similar to the optical spectra, and the IC peak frequencies could be at $\nu_c\sim10^{22}Hz$.
However, the optical index of PKS 0347-211 is slightly larger than the GeV index because the optical data shows a curved spectrum. For PKS 0227-369, the optical spectrum is also steeper than the GeV spectrum due to large data error in infrared band.
For the 4 sources (PKS 0208-512, 3C 273, B2 1520+31, and PKS 0347-211), their GeV spectrum may be of a power-law shape in low energies and show a break at the GeV highest energy end. The spectral break for these sources may be caused by the high-energy cutoff in the electron distribution when the emission region is outside the BLR.
For PKS 0347-211, we can reproduce the optical and  GeV spectrum by adjusting the electron high-energy cutoff, although the optical spectra is slightly steeper than the GeV spectra. The optical and  GeV spectrum may reflect an intrinsic break in the electron distribution for this source.

We know that the mean energy of the IC scattering is given by
\begin{equation}
\epsilon\simeq\frac{3}{2}\gamma'^2\delta^2_{D}\epsilon_{0},
\end{equation}
where $\delta_{D}$ is the Doppler factor of the jet, $\gamma'$ is the Lorentz factor of electron, and $\epsilon_{0}$ is the energy of soft photon (unit of $m_ec^2)$.
The condition for the Thomson scattering is
\begin{equation}
4\gamma'\epsilon_{0}\delta_{D}\lesssim 1.
\end{equation}
Combining above equations, we get a critical energy $\epsilon_c\simeq\frac{3}{32\epsilon_0}$, which is not sensitive to the Doppler factor and only depends on the energy of soft photon. The energy of soft photon from the BLR is $\epsilon_{0}\simeq \epsilon_{Ly{\alpha}}=2\times10^{-5}$, we can obtain the critical frequency $\nu_c\simeq6\times10^{23}$Hz ($E_c\sim2.4$GeV), which is very close to the broken energy in 5 FSRQs. The broken spectrum could be caused by the KN effect, implying that the GeV emission is produced within the BLR. The energy of soft photon from the dust torus is $\nu_{0}\simeq 1\times10^{13}$ Hz, we can obtain the critical frequency $\nu_c\simeq5\times10^{25}$Hz, in which the IC scattering is in the Thomson regime and produces a power-law spectrum below the critical energy. We infer that the GeV emission could be produced outside the BLR for 16 FSRQs. In the next section, we present the detail model to produce the quasi-simultaneous SEDs of 21 FRSQs for testing our method.

\section{Model}
We consider a spherical blob of the jet moving with Lorentz factor $\Gamma$ at a small angle $\theta$ with the line of sight, which is filled by relativistic electrons with a uniform magnetic field. The radiation from the emission region is produced by both the synchrotron radiation and the IC scattering of relativistic electrons. The relativistic effects are described by the Doppler factor $\delta_{D}\simeq\Gamma$, where we assume a view angle of $\theta\simeq 1/\Gamma$. The electron spectrum is described by a broken power-law distribution with the form
\begin{equation}
N(\gamma')=\left\{
    \begin{array}{ll}
    k\gamma'^{-p_{1}}    &\mbox{$\gamma'\leq\gamma'_{b}$}\\
    k{\gamma'_{b}}^{p_{2}-p_{1}}\gamma'^{-p_{2}}   &\mbox{$\gamma'>\gamma'_{b}$},\\
    \end{array}
\right.
\end{equation}
where $\gamma'_{b}$ is broken Lorentz factor, $p_{1,2}$ are the indices of electron distribution below and above $\gamma'_{b}$, and
$\alpha_{1,2}=(p_{1,2}-1)/2$ are the observed spectral indices below and above the synchrotron peak.
Throughout the paper, the unprimed quantities refer to the observer's frame, while the primed ones to the blob's frame.
Photon and electron energies are expressed in reduced units of $m_e c^2$.

The synchrotron flux $\nu F_{\nu}$ is given by \citep{sau04,fin08}
\begin{equation}
f_{\epsilon}^{syn}=\frac{\sqrt{3}\delta_{D}^{4}\epsilon'e^{3}B}{4\pi hd_{L}^2}\int_{1}^{\infty}d\gamma'N(\gamma')R(x),
\end{equation}
where $x=\frac{2\pi\epsilon' m_e^2c^3}{3eBh\gamma'^2}$,
$R(x)=2x^2\{K_{4/3}(x)K_{1/3}(x)-0.6x[K_{4/3}^2(x)-K_{1/3}^2(x)]\}$,
$e$ is the electron charge, $B$ is the magnetic strength, $h$ is the Planck constant, $K_{\alpha}$ is the modified $\alpha$-order Bessel function, $d_{L}$ is the luminosity distance of the source with the redshift $z$, and $\epsilon=\delta_{D}\epsilon'/(1+z)$ is the dimensionless synchrotron photon energy in the observer's frame.

The SSC flux $\nu F_{\nu}$ is given by \citep{fin08}
\begin{equation}
f_{\epsilon_{s}}^{ssc}=\frac{9\sigma_{T}\epsilon'{_s^2}}{16\pi R'{_b^2}}\int_{0}^{\infty}d\epsilon'\frac{f_\epsilon^{syn}}{\epsilon'^{3}}\int_{\gamma'_{min}}^{\gamma'_{max}}
d\gamma'\frac{N(\gamma')}{\gamma'^{2}}F(q,\Gamma_e),
\end{equation}
where $\sigma_{T}$ is the Thomson cross-section, $R'_b$ is the radius of the emitting blob, and $\epsilon_s=\delta_{D}\epsilon_s'/(1+z)$ is the dimensionless scattered photon energy in the observer's frame. The function $F(q,\Gamma_e)$ is given by
$F(q,\Gamma_e)=[2qlnq+(1+2q)(1-q)\nonumber\\+\frac{1}{2}\frac{(\Gamma_e q)^2}{(1+\Gamma_e q)}(1-q)](\frac{1}{4\gamma'} \leq q \leq 1)$, where
$q=\frac{\epsilon'_s/\gamma'}{\Gamma_e(2-\epsilon'_s/\gamma')}$ and $\Gamma_e=4\epsilon'\gamma'$.

In the EC mechanism, the soft photons for the IC scattering include accretion disk photons, reprocessed disk photons by the BLR cloud and IR photons from a dust torus. When the GeV emission region is within the BLR, we consider the photons from accretion disk and the BLR cloud. When  the GeV emission region is outside the BLR, we consider the photons from the dust torus. The thermal emission from the accretion disk is calculated by assuming a standard thin disk \citep{sha73}. The EC emission from the accretion disc is calculated by using the formulation given
by \citet{ghi09}.
We assume that the external radiation field from the BLR or dust torus is characterized by an isotropic blackbody with temperature $T=h\nu_{peak}/(3.93k_B)$, where $\nu_{peak}$ is the peak frequency of this component in a $\nu F_{\nu}$ plot. The $\nu F_{\nu}$ spectrum of the EC process is given by \citep{mar03,fin12}
\begin{eqnarray}
f_{\epsilon}^{EC}=\frac{3c\sigma_{T}\epsilon'{_s^2}\delta_D^4}
{16\pi d_L^2}\frac{15u_{ext}'}{(\pi\Theta')^4}\int_{0}^{\infty}d\epsilon'\frac{\epsilon'}
{exp(\epsilon'/\Theta)-1}\int_{\gamma'_{min}}^{\gamma'_{max}} d\gamma'   \nonumber \\
\times\frac{N(\gamma')}{\gamma'^{2}}F(q,\Gamma_e),
\end{eqnarray}
where $\Theta'=k_BT'/m_ec^2$ is dimensionless temperature, and $k_B$ is the Boltzmann constant. In the comoving frame, the soft photon energy is shifted to $\epsilon'_{peak}\simeq\Gamma \epsilon_{peak}$. The external photon energy density is given by $u_{ext}'\simeq\Gamma^2u_{ext}$.

For the BLR cloud, the photon energy density is $u_{ext,BLR}=\xi_{BLR}L_{d}/(4\pi r_{BLR}^2c)$, in which $L_{d}$ is the accretion luminosity and  $\xi_{BLR}\sim0.1$ is the fraction of $L_d$ reprocessed by the BLR cloud.  The reverberation mapping indicated that the typical size of the BLR
is $r_{BLR}\simeq(1-3)\times10^{17}\sqrt{\frac{L_d}{10^{45}erg\cdot s^{-1}}} cm$ \citep{kas07,ben09}.
The SED of the BLR can be well approximated as an isotropic blackbody with a peak frequency of $\nu_{peak}\simeq1.5\nu_{Ly\alpha}\simeq3.7\times10^{15}Hz $ \citep{tav08}. $r_{BLR}$ is adjusted to agree within a factor of 2 with the value given by the reverberation mapping in our SED modelling.
For the dust torus, its SED is described as an isotropic blackbody with the temperature from $T=200$ K \citep{lan10} to $1200$ K \citep{mal11}.
The inner radius of the dust torus is $r_{IR}\simeq2.5\times10^{18}\sqrt{\frac{L_d}{10^{45}erg\cdot s^{-1}}} cm$ \citep{ghi08,nen08}. The total IR luminosity of the dust is $L_{IR}\simeq4\pi r_{IR}^2\sigma_{SB} T^4$, where $\sigma_{SB}$ is the Stefan-Boltzmann constant. The covering factor of the IR cloud can be estimated as $L_{IR}/L_d$. We can estimate the location of the emission region from the black hole using the covering factor as $r\simeq0.5(\frac{L_d}{L_{IR}}r_{IR}^2)^{1/2}$ \citep{kus13}.

For a broken electron distribution, there is $F_{\nu}\propto\nu^{\alpha_2}$ above the synchrotron peak frequency. When the IC scattering is in the Thomson regime, there is $F_{\nu}\propto\nu^{\alpha_2}$ above the IC peak frequency, implying that the synchrotron and IC spectra above the peak frequencies have the same spectral index. However, when the IC scattering takes place at the KN regime, the synchrotron and IC spectra above the peak frequencies have  different shapes. We try to constrain the location of the GeV emission by the spectral shapes.

\section{Application}
We use the model described in Section 2 to reproduce the quasi-simultaneous SEDs of 21 FRSQs for inferring the location of the GeV emission, in which the observed data of these 21 FSRQs are taken from \citet{abd10b}.  We report the observed minimum variability time-scale when they are available from
the literature. The observed minimum variability time-scale constrains the size of the emission region via $R'_b\leq\delta_Dct_{var}/(1+z)$. There is a strong blue component in the SEDs of 3C 273 which could be the emission from accretion disk \citep{abd10b}. The thermal emission from accretion disk is generally swamped by the strong beaming jet emission. There are no prominent blue components in UV energies for the other sources. The accretion disk luminosity and the mass of the black hole are either obtained from the literature or derived through the SED modelling. The modelling parameters are listed in Table 2-5 for 21 FSRQs.
It can be found that when the emission is inside the BLR, the magnetic field  $B$ is about a few G and the energy density $u_{BLR}$ is about $10^{-2}-10^{-3} erg\cdot cm^{-3}$, and the location of the emission from the central engine is $r\lesssim0.1$ pc.
When the emission is outside the BLR, $B\sim0.1-1$ G,  $u_{IR}\sim10^{-4}-10^{-6} erg\cdot cm^{-3}$,
the location of the emission is at a few parsecs up to tens of parsecs from the central engine, and the temperature of dust is $T\sim200-700$ K. The previous studies also suggested the presence of a warm dust with the temperature $150-800$ extension up to a few tens of parsec \citep{jaf04,cle07,lan10}.
In both cases, the Doppler factor $\delta_D$ is from 15 to 28 which is consistent with ones measured by VLBI for FSRQs \citep{jor05}. The $\gamma'_b$ is clustered at $1\times10^3$, which is consistent with the values given by \citet{ghi98} and \citet{ghi02}. The electron spectral index below the broken energy is $p_1\sim2$, which is expected by the fast cooling regime for the FSRQs \citep{ghi02,fin13}.
The size of the emitting region from the SED modelling corresponds to the variability time-scale of a few days, consistent with the day-scale variability seen in most Fermi-detected FSRQs. Our results suggest that the physical parameters in the emission strongly depend on the location of the emission region. It is obvious that the magnetic field, the size of the emission region and the energy density are higher when the emission region is nearer to the black hole. On the other hand, the shapes of the $\gamma$-ray spectra are also significantly different when the $\gamma$-ray region is inside or outside the BLR.

The equipartition parameter between the comoving electron and magnetic field energy density is calculated as $\kappa_{eq}=u'_B/u'_e=\int_{\gamma'_{min}}^{\gamma'_{max}}\gamma'm_ec^2n'(\gamma')d\gamma'/(B^2/8\pi)$, where $n'(\gamma')=\frac{3N(\gamma')}{4\pi {R'_b}^3}$ is electron number density per volume. Our results show that the ratio of the magnetic field and electron energy density is near equipartition for all sources, which is consistent with the SED modelling of FSRQs, typically $0.1\leq\kappa_{eq}\leq1$.
In fact, if the X-rays of FRSQs are produced by the SSC process, the $\gamma'_b$ will not be large and can be estimated by \begin{equation}
\gamma'_b\simeq 10^3 \left(\frac{\nu_c}{10^{17}Hz}\right)^{-\frac{1}{2}}\left(\frac{\nu_s}{10^{11}Hz}\right)^{-\frac{1}{2}},
\end{equation}
where $\nu_c$ and $\nu_s$ are the SSC and synchrotron photons corresponding to X-ray and radio bands. It is shown that the electrons could not be accelerated to the high energy due to the fast cooling by strong external photon field.  \citet{ghi10} reproduce the SEDs of the 57 FSRQs assuming the emission region locating at the $0.01-0.5$ pc from the central black hole, however  most of the multi-wavelength data they used are not simultaneous or quasi-simultaneous. We reproduce the quasi-simultaneous SEDs of 21 FRSQs using the one-zone leptonic model, which is shown in Figure 1-2. It is shown that the SSC model can't explain the SEDs of the FSRQs alone, implying that the GeV emission originates from the EC process, and the X-ray emission is produced by the SSC process.

The quasi-simultaneous SEDs of 5 FRSQs are reproduced by the SSC and IC scattering of photons from the disk and BLR  by the electrons, which is shown in Figure 1. It can be seen that the GeV emission comes from the combination of the Compton-scattered disk and BLR radiations.
Besides, the observed GeV spectrum steeps at a few GeV energy which is near the transition energy from  the Thomson to the KN regime.
It is obvious that the broken spectrum at the GeV bands is caused by the KN effect for 5 FRSQs.
We can't obtain an acceptable fit to the observed GeV spectrum using a dust torus model. On the one hand, the optical and GeV spectrum have
different shapes. On the other hand, the photon energy from the dust torus is too low to produce $\gamma$-ray with the broken spectrum when $\gamma'_b\sim10^3$. It is noted that the soft photon energy from the BLR is 100 times larger than one from the dust torus. The threshold for $\gamma-\gamma$ absorption from the BLR radiation implies that the $\gamma$-rays with energy $\epsilon_1>\frac{2}{\epsilon_{Ly{\alpha}}}\simeq 1\times10^{5}$ ($\nu_1>1\times10^{25}$Hz) will be absorbed \citep{fin10}, in which the energy is above the highest energy photon observed with the Fermi-LAT in our sources except for PKS1502+106 whose highest energy photon is at $\nu\simeq3\times10^{25}$Hz. However, the IC scattering could't reach such high energy due to the KN effect in our model, we will neglect the absorption from the BLR.
The quasi-simultaneous SEDs of 16 FRSQs are reproduced by the SSC and IC scattering of photons from the dust torus by the electrons, which is shown in Figure 2. It can be seen that the observed GeV spectrum has a similarly power-law shape and the IC peak frequency is at $\nu_c\sim10^{22}$Hz.
We can obtain an acceptable fit to the observed GeV spectrum using a dust torus model, because the IC scattering takes place at the Thomson regime, which produce a power-law spectrum below the high energy spectrum cut-off  and the IC peak energy is very close to the peak energy of SED.
The 4 sources out of 16 FRSQs show a spectral break at the GeV highest end.
We can produce the broken spectrum by adjusting the high-energy electron cutoff, $\gamma_{max}$. We suggest that the spectral break is caused by
an intrinsic break in the electron distribution.
However, we can't provide an acceptable fit to the observed SED using a BLR model, because the photon energy from the BLR is too high to produce an IC spectrum with peak energy at $\sim 10^{22}Hz$ when $\gamma'_b\sim10^3$. We can estimate the frequency of the soft photon by the IC peak frequency, given by $\nu_0\simeq3\times10^{13}(\frac{\nu_c}{10^{22}Hz})(\frac{\delta_D}{15})^{-2}(\frac{\gamma'_b}{10^3})^{-2} Hz$, which is consistent with the frequency from the dust torus emission.

The quasi-simultaneous optical-IR and $\gamma$-ray data are important in the test of our method. In the Thomson regime, the $\gamma$-ray spectrum should have the same spectral index as the optical-IR spectrum. But in the KN regime, the $\gamma$-ray spectrum should be softer than the optical-IR spectrum. However, there are only a few optical data for many sources, future observations in optical-IR bands are needed. We only infer that the GeV emission is produced within the BLR for 5 FSRQs and outside the BLR for 16 FSRQs based on the quasi-simultaneous data from \citet{abd10b}. The mean IC peak frequency is $<\nu_c>\simeq10^{22}$Hz for the most powerful blazars \citep{gio12}, implying that the IC scattering usually is in the Thomson regime, and supporting that the GeV emission region is outside the BLR for the most FSRQs.

\section{Conclusion and Summary}
The origin of GeV spectral break is still unknown. \citet{fin10} suggested that a combination of disk and BLR scattering components can produce the
spectral break, but this solution requires a BLR with a wind-like profile.
\citet{pou10} suggested that the spectral break is caused by the absorption from the BLR, however, this explanation has been brought into question
\citep{har12}. Recently, \citet{cer13} suggested that the  spectrum break of 3C 454.3 is produced by the KN effect, as has been proposed to explain
the SED of PKS 1510-089 \citep{abd10c}.
The KN effect due to the up-scattered BLR radiation would produce a broken spectrum at a few GeV energies. Since the broken energy is insensitive to the Doppler factor and only depend on the energy of soft photon, the broken energy approximately keeps constant due to the  KN effect.
The observed broken spectrum doesn't significantly vary with the flux states, it is very likely that the broken spectrum is caused by the KN effect.
We suggest that the broken spectrum of 5 FSRQs is caused by the KN effect.
To date, three FSRQs have been detected at the VHE $\gamma$-ray \citep{lin13}. The VHE photons are severely attenuated by the BLR photon through pair production if the emission region is located inside the BLR \citep{li06,sit08}. However, if the emission region is located far beyond the BLR, the VHE photons can avoid the severe $\gamma$-$\gamma$ absorption and the KN effect has a weak effect on the VHE spectrum until the energy is far larger than $10^{25}Hz$.
Furthermore, \citet{abd09b} suggested that the spectral break may be caused by the high-energy cut-off in the electron distribution when the emission  region is outside the BLR. This is also in accordance with our results for 4 FSRQs.
We suggested that the GeV emission from 16 FSRQs is produced by the IC scattering of the IR photons from the dust.

We try to infer the location of the GeV emission for 21 FSRQs with quasi-simultaneous SEDs based on the spectral shapes, in which the SEDs of 21 FSRQs are reproduced by the one-zone leptonic model including the SSC and EC processes. Our results suggest that the X-ray emission could be produced by the SSC process and the GeV emission could originate from the EC process, in which the EC emission originates from the IC scattering of photons from the disk and BLR or dust torus by electrons. When the GeV emission is produced within the BLR, the IC scattering could occur at the KN regime and produce a broken spectrum at the GeV bands. When the GeV emission is produced outside the BLR, the IC scattering could take place at the Thomson regime and produce a GeV spectrum similar with electron distribution. We infer that the GeV emission region of 5 FSRQs and 16 FRSQs is located within and outside the BLR  based on the quasi-simultaneous data, respectively. In this case, we find that the magnetic field and electron energy density are near equipartition.

\section*{Acknowledgments}
We thank the anonymous referee for valuable comments and suggestions.
We acknowledge the financial supports from the National Basic Research Program of China
(973 Program 2009CB824800), the National Natural Science Foundation
of China 11133006, 11163006, 11173054, and the Policy Research Program of Chinese Academy of
Sciences (KJCX2-YW-T24).
%--------------------------------------------------------------------------------------------------------------------------------------------------------

\bibliography{refernces}

%--------------------------------------------------------------------------------------------------------------------------------------------------------
\clearpage

\begin{table*}
\begin{minipage}{160mm}
\caption{A simple power-law or broken power-law fit for the spectra of 21 FSRQs. $\alpha_{GeV,1}$ and $\alpha_{GeV,2}$ are the energy spectral index of the GeV spectrum from the broken power-law fit below and above the broken energy $E_b$. $\alpha_{GeV}$ is the energy spectral index of the GeV spectrum from the power-law fit,
$\alpha_{Opt}$ is the energy spectral index of the optical spectrum from the power-law fit}
\begin{tabular}{lcccccccccccccc}
\hline

Name     &$\alpha_{GeV}$  &$\chi^2$/d.o.f. &$\alpha_{GeV,1}$ &$\alpha_{GeV,2}$ &$E_b$   &$\chi^2$/d.o.f.   &$\alpha_{Opt}$   &$\chi^2$/d.o.f.\\
S4 0133+47   &$1.04\pm0.09$ &0.19 &               &             &                &      &$1.61\pm0.06$       &1.27    \\
4C 66.20     &              &     &$1.03\pm0.08$ &$1.86\pm0.81$ &$3.52\pm2.72$   &0.25                                \\
PKS 0454-234 &              &     &$1.12\pm0.04$ &$1.74\pm0.45$ &$3.51\pm2.15$   &0.03  &$1.70\pm0.07$       &1.19    \\
PKS 1502+106 &              &     &$1.07\pm0.03$ &$1.78\pm0.21$ &$4.43\pm1.00$   &2.12  &$1.76\pm0.07$       &1.26    \\
PKS 1454-354 &              &     &$0.98\pm0.08$ &$1.48\pm0.16$ &$0.92\pm0.80$   &0.01  &$1.85\pm0.15$       &1.23    \\
\hline
PKS 2325+093 &$1.52\pm0.10$ &0.56  \\
S3 2141+17   &$1.28\pm0.07$ &0.13  \\
OT 081      &$1.25\pm0.07$ &0.27 &               &             &                &      &$1.35\pm0.14$       &0.15    \\
PKS 0208-512 &$1.23\pm0.05$ &1.22 &               &             &                &      &$1.13\pm0.12$       &3.30    \\

PKS 0528+134 &$1.48\pm0.08$ &0.55 &               &             &                &      &$1.40\pm0.16$       &1.08    \\
1Jy 1308+326 &$1.21\pm0.06$ &0.23  \\
3C 279       &$1.25\pm0.05$ &0.61 &               &             &                &      &$1.32\pm0.16$       &1.92    \\
3C 273       &$1.61\pm0.04$ &3.30 &               &             &                &      &$1.57\pm0.32$       &0.11    \\

4C 29.45     &$1.50\pm0.11$ &0.53  \\
PKS 0727-11  &$1.26\pm0.05$ &1.65  \\
S4 0917+44   &$1.30\pm0.09$ &0.73  \\
B2 1520+31  &$1.32\pm0.05$ &2.75 &               &             &                &      &$1.46\pm0.22$       &1.61    \\

PKS 0420-01  &$1.57\pm0.09$ &2.34 &               &             &                &      &$1.43\pm0.12$       &0.15    \\
PKS 0347-211 &$1.46\pm0.08$ &0.78 &               &             &                &      &$1.84\pm0.29$       &0.58    \\
4C 28.07     &$1.32\pm0.13$ &0.25  \\
PKS 0227-369 &$1.52\pm0.09$ &0.33 &               &             &                &      &$1.88\pm0.42$       &0.11    \\
\hline
\end{tabular}
\end{minipage}
\end{table*}

\begin{table*}
\begin{minipage}{160mm}
\caption{Model parameters used in our SED modelling for 5 FSRQs.}
\begin{tabular}{lccccccccccccccc}
\hline
Name  &z &$B(G)$ &$\delta_D$ &$\gamma'_{min}$ &$\gamma'_{b}$ &$\gamma'_{max}$ &$p_1$ &$p_2$  &$R'_b(cm)$ &$t_{var,min}$ \\
~[1]      &[2] &[3] &[4] &[5] &[6] &[7] &[8] &[9] &[10] &[11]\\
\hline
S4 0133+47   &0.859 &2.50  &28 &1E2   &1.3E3 &2E4 &2   &4.4  &4.0E15   &2.5 hr\\
4C 66.20     &0.657 &1.24  &20 &1E2   &2.0E3 &3E4 &2   &4.0  &5.5E15   &4.2 hr\\
PKS 0454-234 &1.003 &1.00  &25 &1E2   &1.8E3 &1E5 &2   &4.5  &8.8E15   &6.5 hr\\
PKS 1502+106 &1.839 &1.15  &28 &1E2   &2.2E3 &1E5 &1.8 &4.5  &6.8E15   &6.4 hr\\
PKS 1454-354 &1.424 &1.70  &22 &1E2   &1.7E3 &1E5 &2   &4.5  &7.5E15   &7.7 hr\\
\hline
\end{tabular}
\quad\quad\quad\quad\quad\quad\quad
Note. Columns 11 is the expected minimum variability time-scale from the SED modelling.
\end{minipage}
\end{table*}

\begin{table*}
\begin{minipage}{160mm}
\caption{Model parameters used in our SED modelling for 5 FSRQs.}
\begin{tabular}{lcccccccccccccccccccccccccc}
\hline
Name   &$\nu_{peak}(Hz)$ &$u_{BLR} (erg\cdot cm^{-3})$ &$r (cm)$ &$\kappa_{eq}$  &$L_d (erg\cdot s^{-1})$  &$M_9$ &Ref
&$t_{var}$ &Ref \\
~[1]      &[2] &[3] &[4] &[5] &[6] &[7] &[8] &[9] &[10]  \\
\hline
$S4 0133+47$ &3.75E15 &1.73E-3 &2.4E17   &0.88   &2.75E45   &1.60  &Gu01    &\ldots &\ldots \\
4C 66.20     &3.75E15 &2.08E-3 &3.0E17   &0.17   &2.63E45   &4.15  &Gu01    &\ldots &\ldots \\
PKS 0454-234 &3.75E15 &3.69E-3 &6.9E16   &0.34   &2.57E45   &0.11  &Fan09   &18 hr  &Vov13 \\
PKS 1502+106 &3.75E15 &1.08E-2 &1.4E17   &0.15   &1.10E46   &0.55  &Liu06   &12 hr  &Abd10a \\
PKS 1454-354 &3.75E15 &1.00E-2 &2.0E17   &0.39   &1.00E46   &2     &\ldots  &12 hr  &Abd09a  \\
\hline
\end{tabular}
Note. Column 4 is the location of the emission region from the central engine by the SED modelling. Columns 7 and 8 are the black hole masses in the unit of $10^{9}M_{\odot}$ and the references. Columns 9 and 10 are the measured minimum variability time-scale and the references.
References:
Gu01:  Gu et al. (2001);
Liu06: Liu et al. (2006);
Fan09: Fan et al. (2009);
Abd09a: Abdo et al. (2009a);
Abd10a: Abdo et al. (2010a); and
Vov13: Vovk \& Neronov (2013);
\end{minipage}
\end{table*}

\begin{table*}
\begin{minipage}{160mm}
\caption{Model parameters used in our SED modelling for 16 FSRQs.}
\begin{tabular}{lccccccccccccccc}
\hline
Name  &z &B(G) &$\delta_D$ &$\gamma'_{min}$ &$\gamma'_{b}$ &$\gamma'_{max}$ &$p_1$ &$p_2$  &$R'_b$(cm) &$t_{var,min}$ \\
\hline
PKS 2325+093 &1.843 &0.43   &18  &2.0E2   &1.5E3 &5.0E4  &1.8   &4     &5.5E16     &3.4 d\\
S3 2141+17   &0.213 &0.49   &23  &1.0E2   &1.2E3 &3.4E4  &2     &3.5   &2.4E16     &12 hr\\
OT 081       &0.322 &0.33   &15  &1.0E2   &1.0E3 &1.0E5  &2     &3.5   &3.5E16     &1.2 d\\
PKS 0208-512 &1.003 &0.28   &15  &1.0E2   &1.2E3 &2.5E4  &2     &3.4   &7.0E16     &3.4 d\\

PKS 0528+134 &2.070 &0.45 &19  &3.0E2   &1.2E3 &5.0E4 &1.8   &4.0  &3.0E16   &1.9 d\\
1Jy 1308+326 &0.997 &0.36 &15  &1.0E2   &1.4E3 &4.0E4 &2     &3.4  &4.0E16   &2.1 d\\
3C 279       &0.536 &0.35 &15  &1.0E2   &1.0E3 &4.0E4 &2     &3.4  &3.7E16   &1.5 d\\
3C 273       &0.158 &0.50 &10  &1.0E2   &1.0E3 &8.0E3 &2     &4.1  &2.3E16   &14 hr\\

4C 29.45     &0.729 &0.38  &15 &1.0E2   &1.8E3 &5.0E4 &2   &4.0  &6.0E16   &2.7 d\\
PKS 0727-11  &1.589 &0.27  &18 &1.0E2   &1.5E3 &1.0E5 &2   &3.7  &6.4E16   &3.5 d\\
B2 1520+31   &1.487 &0.21  &18 &1.0E2   &1.1E3 &2.0E4 &2   &3.5  &5.4E16   &2.9 d\\
S4 0917+44   &2.190 &0.85  &23 &1.0E2   &1.0E3 &5.0E4 &2   &3.6  &2.5E16   &1.3 d\\

PKS 0420-01  &0.916 &0.40 &15 &1.0E2   &1.5E3 &4.0E4 &2   &4.0  &7.0E16  &3.4d\\
PKS 0347-211 &2.944 &0.58 &20 &1.0E2   &2.8E3 &2.0E4 &2   &3.8  &5.0E16  &3.8d\\
4C 28.07     &1.213 &0.33 &15 &1.0E2   &1.5E3 &8.0E4 &2   &3.7  &4.7E16  &2.7d\\
PKS 0227-369 &2.115 &0.50 &19 &2.0E2   &1.2E3 &8.0E4 &2   &4.0  &6.0E16  &3.8d\\
\hline
\end{tabular}
\medskip

\end{minipage}
\end{table*}

\begin{table*}
\begin{minipage}{150mm}
\caption{Model parameters used in our SED modelling for 16 FSRQs.}
\begin{tabular}{lcccccccccccccccccccccccccc}
\hline
Name   &$T (K)$ &$u_{IR} (erg\cdot cm^{-3})$ &$r (cm)$ &$\kappa_{eq}$ &$L_d$  &$M_9$ &Ref  &$t_{var}$ &Ref \\
\hline
PKS 2325+093 &440  &2.91E-4 &1.33E19 &0.35  &1.3E46   &1     &\ldots   &\ldots   &\ldots \\
S3 2141+17   &220  &1.00E-5 &1.22E19 &1.40  &1.0E45   &0.14  &Pal05    &3 d      &Vov13  \\
OT 081       &244  &1.59E-4 &4.45E18 &0.34  &2.0E44   &0.47  &Wu09     &3 d      &Vov13  \\
PKS 0208-512 &240  &2.51E-4 &2.68E19 &0.32  &8.0E45   &0.6   &Fan09    &4 d      &Vov13  \\

PKS 0528+134 &550  &6.23E-4 &9.93E18 &0.36  &2.3E46   &0.8   &Fan09   &2 d      &B\"{o}t98 \\
1Jy 1308+326 &360  &3.46E-4 &6.98E18 &0.14  &2.5E45   &0.87  &Che09   &\ldots   &\ldots    \\
3C 279       &366  &1.77E-4 &6.24E18 &0.13  &2.0E45   &0.07  &Fan09   &2 d      &B\"{o}t07b \\
3C 273       &440  &4.31E-5 &3.49E19 &0.18  &1.3E47   &2.4   &Wag08   &1 d      &Cou88     \\

4C 29.45     &320  &1.32E-4 &1.72E19 &0.55  &3.0E45   &0.36  &Xie05   &4.5 d    &Vov13  \\
PKS 0727-11  &400  &3.80E-4 &9.17E18 &0.14  &5.0E45   &0.10  &\ldots  &\ldots   &\ldots \\
B2 1520+31   &220  &9.19E-4 &1.94E19 &0.15  &2.5E45   &0.20  &\ldots  &\ldots   &\ldots \\
S4 0917+44   &661  &3.34E-4 &1.13E19 &0.45  &5.1E46   &7.85  &Che09   &1.4 d    &Vov13  \\

PKS 0420-01  &380  &1.37E-4 &8.51E18 &0.41  &4.3E45   &0.11  &Fan09   &4 d      &D'Ar07 \\
PKS 0347-211 &250  &8.91E-4 &3.42E19 &1     &1.3E46   &5     &\ldots  &\ldots   &\ldots \\
4C 28.07     &300  &2.48E-4 &8.06E18 &0.14  &1.5E45   &0.10  &Fan09   &\ldots   &\ldots \\
PKS 0227-369 &450  &6.64E-4 &8.01E18 &0.96  &7.5E45   &1     &\ldots  &\ldots   &\ldots \\
\hline
\end{tabular}

%\vskip 0.1mm
References:
Gu01:  Gu et al. (2001);
Pal05: Paltani \& T\"urler (2005);
Xie05: Xie et al. (2005);
Liu06: Liu et al. (2006);
Wag08: Wagner (2008);
Fan09: Fan et al. (2009);
Che09: Chen et al. (2009);
Cou88: Courvoisier et al. (1988)
B\"{o}t98: B\"{o}ttcher et al. (1998);
B\"{o}t07b: B\"{o}ttcher et al. (2007b);
D'Ar07: D'Arcangelo et al. (2007); and
Vov13: Vovk \& Neronov (2013);
\end{minipage}
\end{table*}

\clearpage
\begin{figure*}
\centerline{\epsfig{file=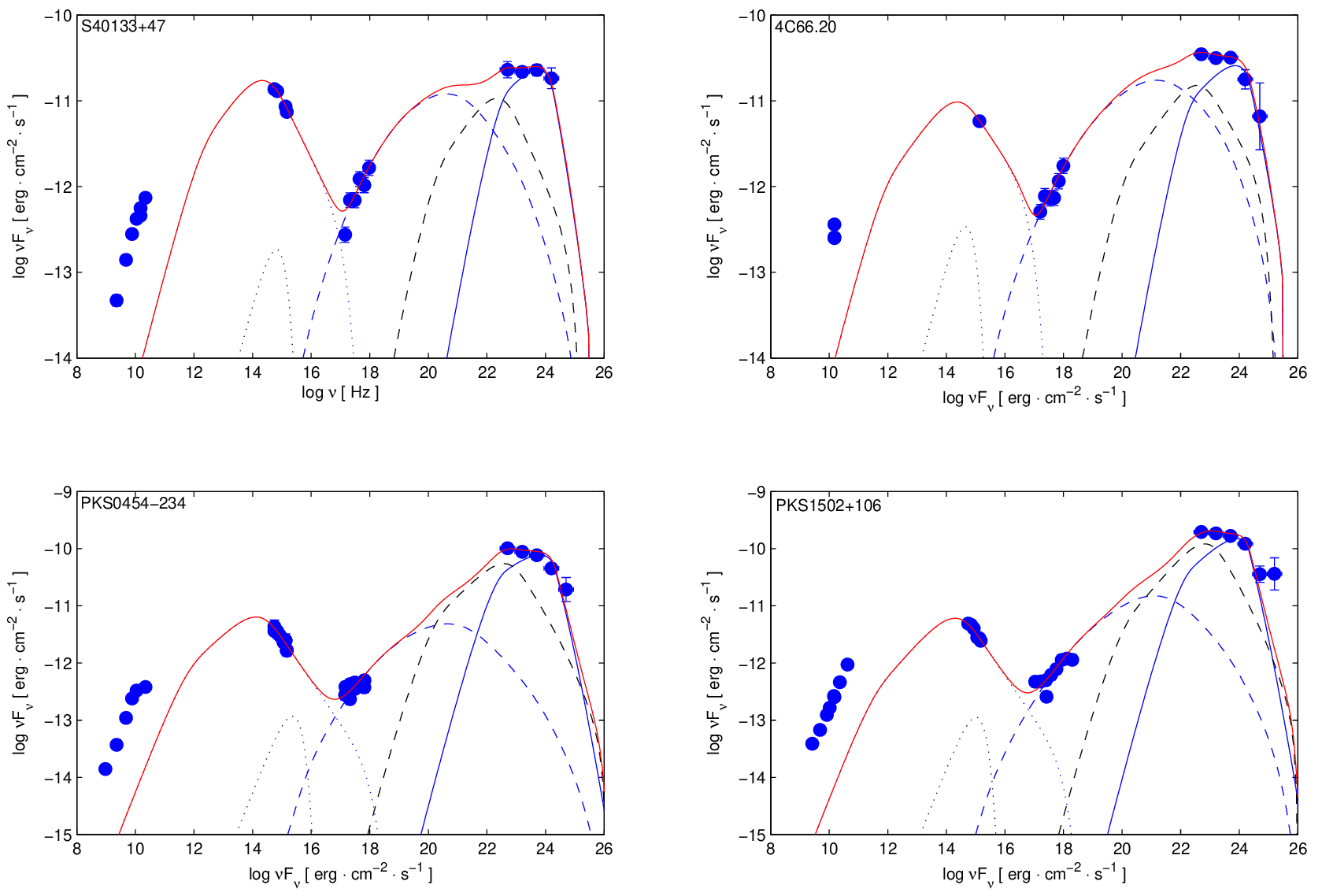, width=15cm}}
\end{figure*}

\begin{figure*}
\epsfig{file=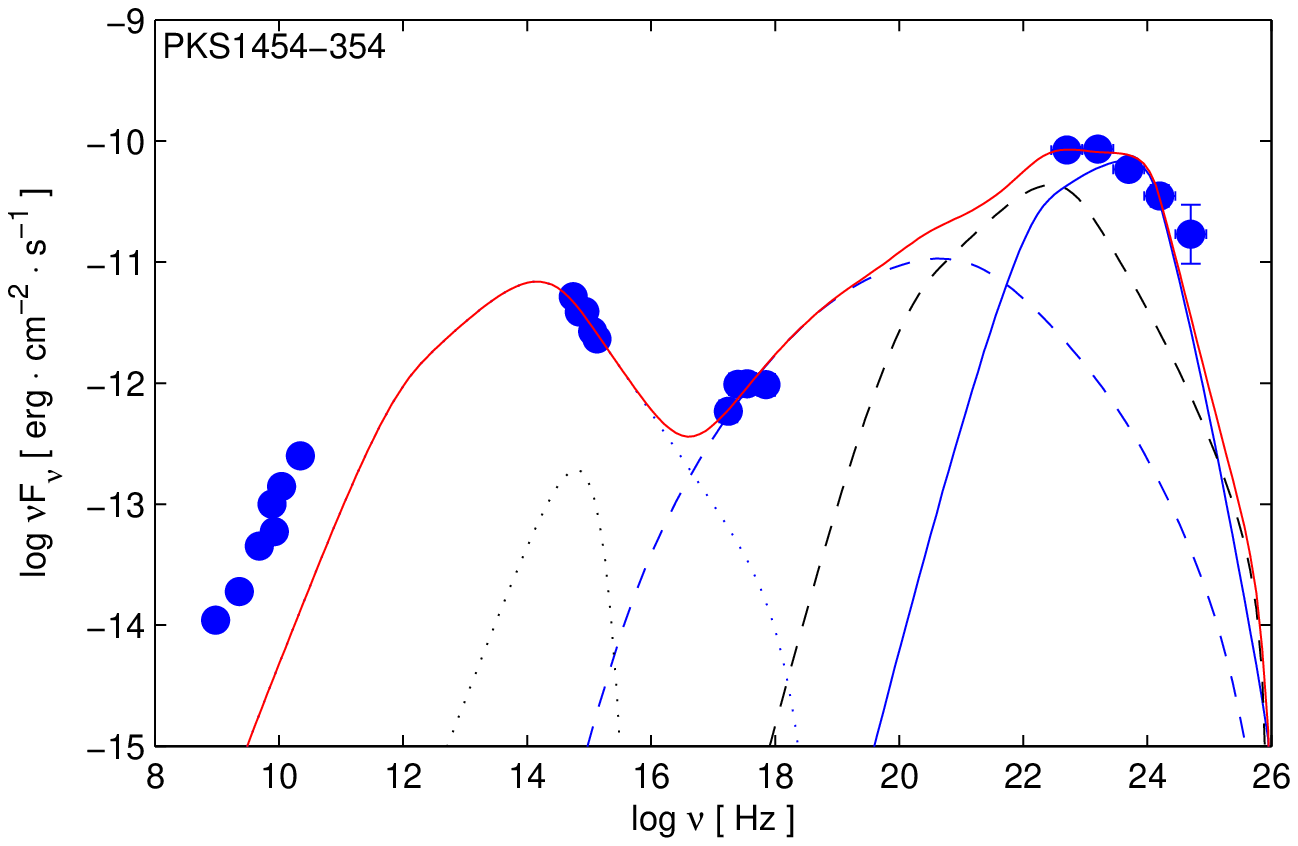, width=7.1cm}
\rightskip = 9.28cm
\caption{The observed SED is shown by solid points with the modelling fits, in which the black dotted line is thermal emission from the accretion disk, the black dashed line is the Compton-scattering disk radiation, the blue dotted, dashed and solid lines are the synchrotron, SSC and the Compton-scattering BLR radiation, respectively. The red solid curve represents the multiwavelength model. \label{fig1}}.
\label{fig2}
\end{figure*}

\clearpage
\begin{figure*}
\centerline{\epsfig{file=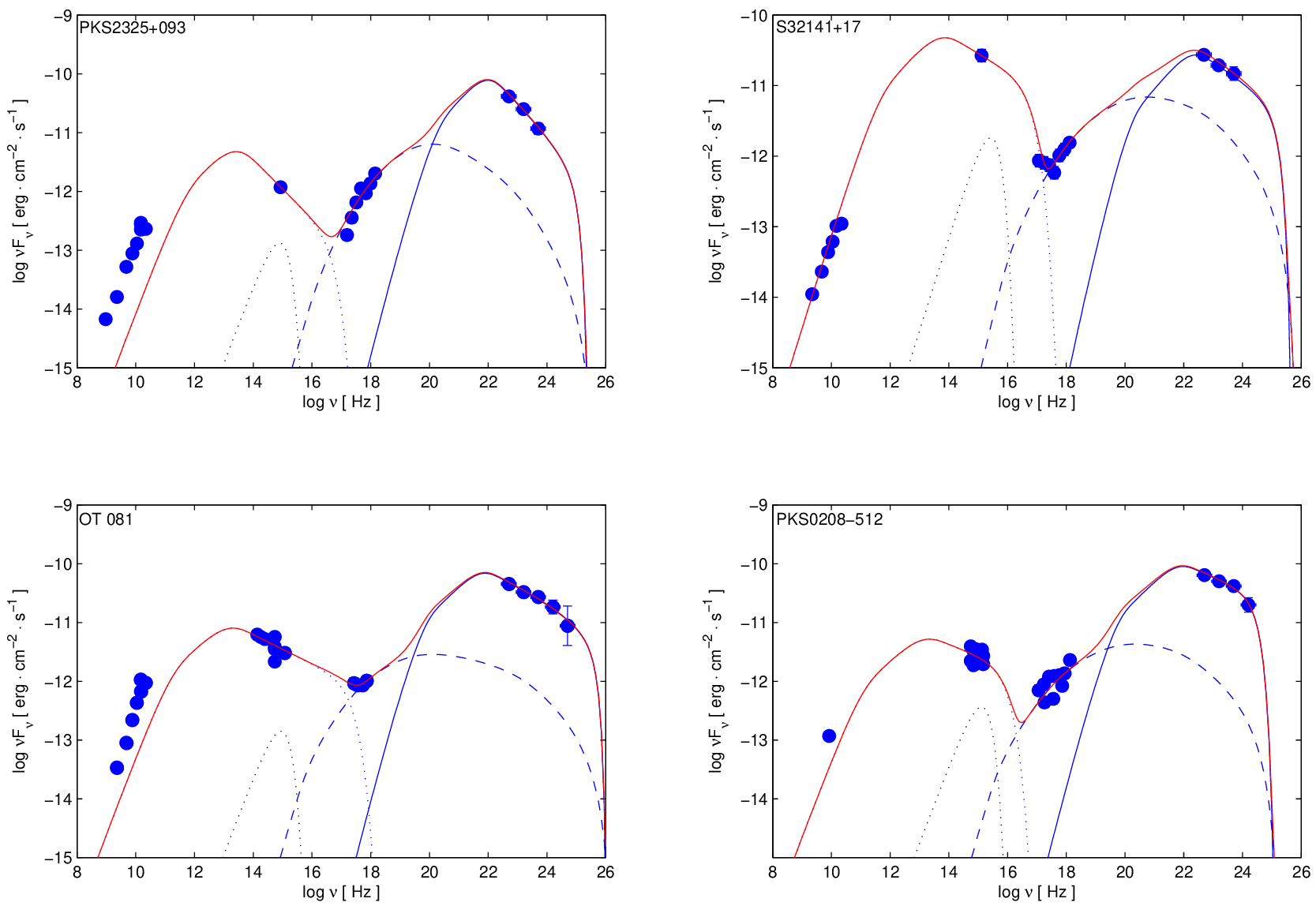, width=15cm}}
\end{figure*}

\begin{figure*}
\centerline{\epsfig{file=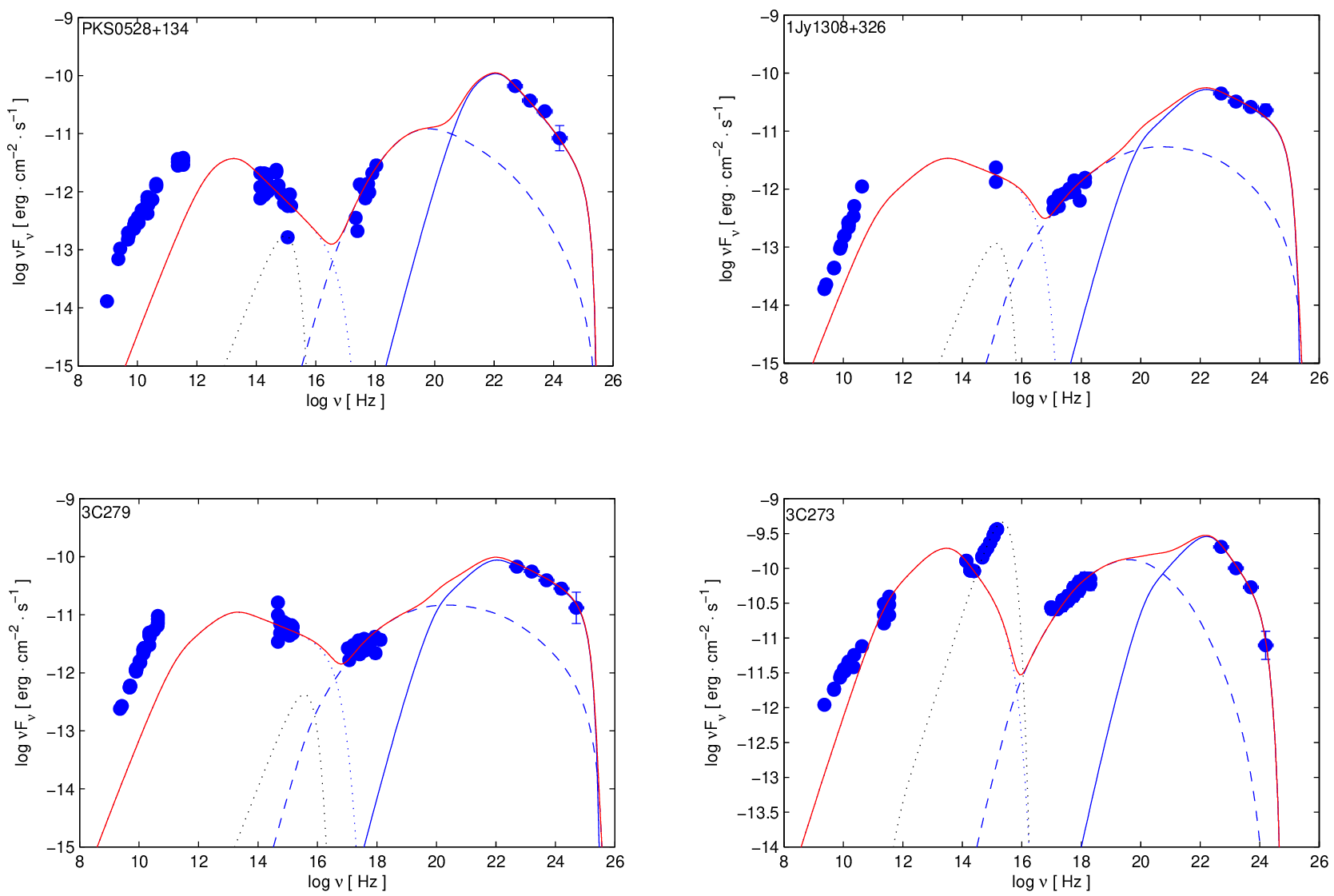, width=15cm}}
\end{figure*}

\begin{figure*}
\centerline{\epsfig{file=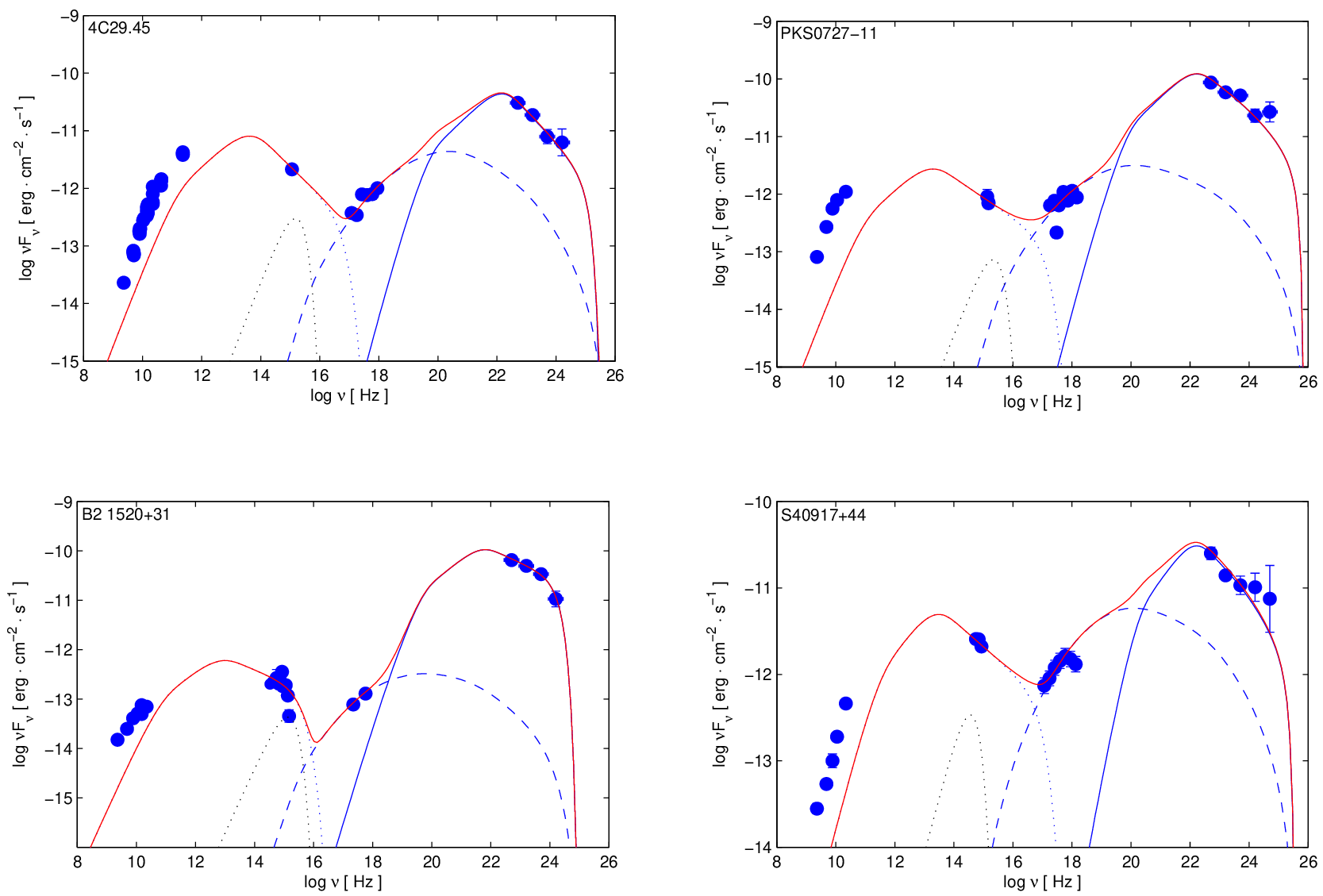, width=15cm}}
\end{figure*}

\begin{figure*}
\centerline{\epsfig{file=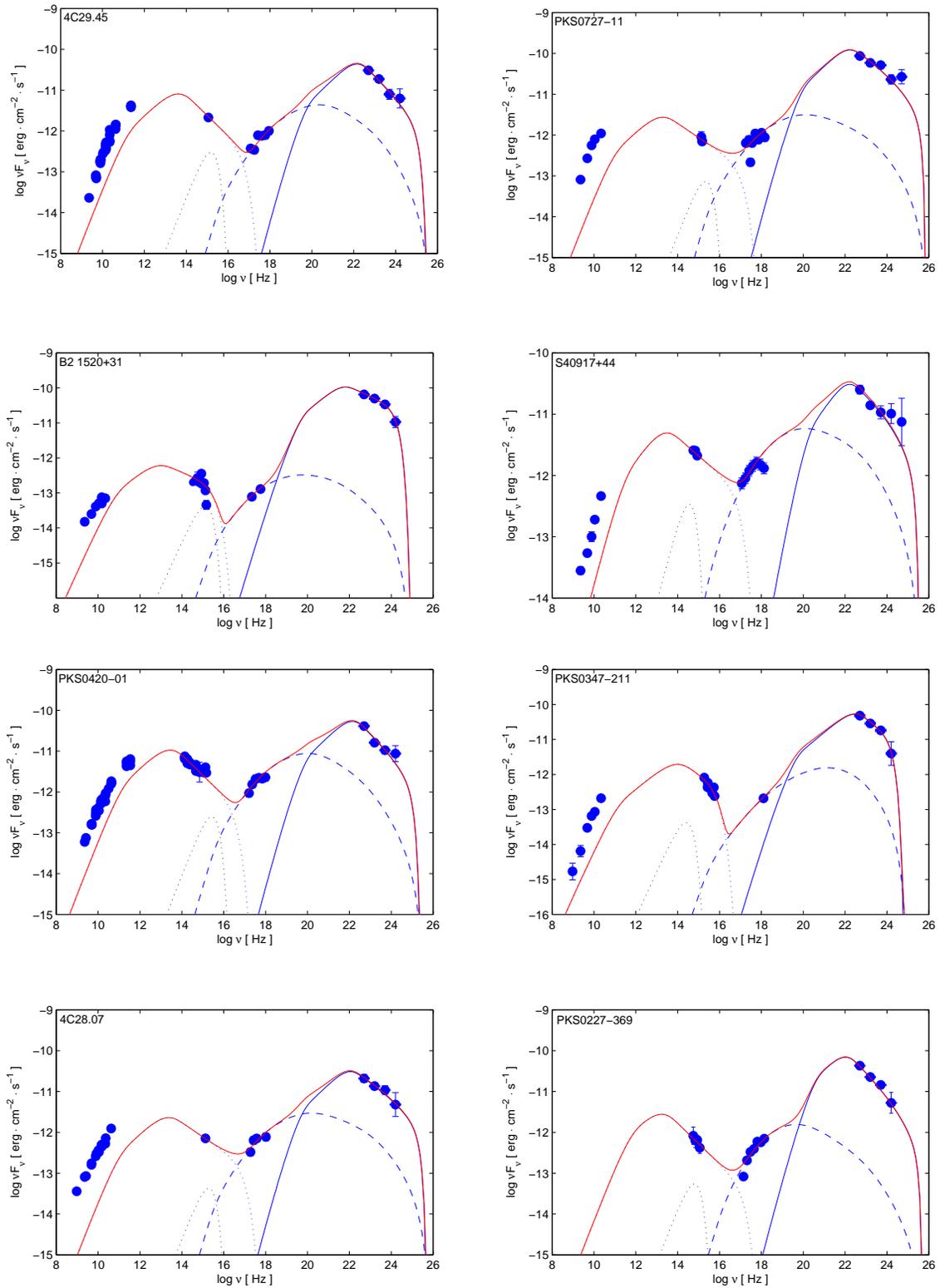, width=15cm}}
\caption{The observed SED is shown by solid points with the modelling fits, in which the black dotted line is thermal emission from the accretion disk, the blue dotted, dashed and solid lines are the synchrotron, SSC and the Compton-scattering dust radiation, respectively. The red solid curve represents the multiwavelength model.}.
\label{fig2}
\end{figure*}

\end{document}